\documentclass[pra,aps,twocolumn,longbibliography,superscriptaddress]{revtex4-1}
\usepackage{graphicx,bm}
\usepackage{amsmath}
\usepackage{txfonts}
\usepackage{color}
\usepackage{hyperref}
\usepackage{mathdots}

\newcommand{\rp}[1]{(\ref{#1})}

\newcommand{\br}[1]{\langle #1|}

\newcommand{\ke}[1]{|#1\rangle}

\newcommand{\da}{^\dagger}

\newcommand{\pt}[1]{\left( #1 \right)}
\newcommand{\pq}[1]{\left[ #1 \right]}
\newcommand{\pg}[1]{\left\{ #1 \right\}}

\newcommand{\bs}[1]{\boldsymbol #1}

\newcommand{\lpg}[1]{\left\{ #1 \right.}

\newcommand{\rpg}[1]{\left. #1 \right\}}
\newcommand{\ee}{{\rm e}}
\newcommand{\ii}{{\rm i}}

\newcommand{\id}{\openone}

\newcommand{\nn}{{\nonumber}}

\newcommand{\mat}[2]{
                      \begin{array}{#1}
                       #2
                       \end{array}  }

\newcommand{\matt}[2]{ \pt{
                      \begin{array}{cc}
                       #1 \\
                       #2
                     \end{array}  }  }

\newcommand{\vb}{{\bf b}}

\newcommand{\vx}{{\bf x}}

\newcommand{\AAA}{{\cal A}}
\newcommand{\BB}{{\cal B}}
\newcommand{\CC}{{\cal C}}
\newcommand{\DD}{{\cal D}}
\newcommand{\GG}{{\cal G}}
\newcommand{\EE}{{\cal E}}

\newcommand{\KK}{{\cal K}}

\newcommand{\MM}{{\cal M}}

\newcommand{\OO}{{\cal O}}
\newcommand{\PP}{{\cal P}}
\newcommand{\QQ}{{\cal Q}}
\newcommand{\RR}{{\cal R}}
\newcommand{\TT}{{\cal T}}
\newcommand{\UU}{{\cal U}}
\newcommand{\VV}{{\cal V}}
\newcommand{\WW}{{\cal W}}
\newcommand{\XX}{{\cal X}}
\newcommand{\YY}{{\cal Y}}
\newcommand{\ZZ}{{\cal Z}}
\newcommand{\SSS}{{\cal S}}

\usepackage[normalem]{ulem}

\begin{document}

\title{
Possibility to generate any Gaussian cluster state by a multi-mode squeezing transformation
}

\author{Stefano Zippilli}
\affiliation{School of Science and Technology, Physics Division, University of Camerino, I-62032 Camerino (MC), Italy}
\author{David Vitali}
\affiliation{School of Science and Technology, Physics Division, University of Camerino, I-62032 Camerino (MC), Italy}
\affiliation{INFN, Sezione di Perugia, I-06123 Perugia, Italy}
\affiliation{CNR-INO, L.go Enrico Fermi 6, I-50125 Firenze, Italy}

\date{\today}

\begin{abstract}
Gaussian cluster states are ideal infinitely squeezed states. In practice it is possible to construct only approximated version of them with finite squeezing. Here we show 
how to determine the specific multi-mode squeezing transformation, which generates a faithful approximation of any given Gaussian cluster state.
\end{abstract}
\maketitle


Cluster states are highly entangled states which are the fundamental resource for measurement based quantum computation~\cite{raussendorf2001,menicucci2006}. In the continuous variable setting the interest in
Gaussian cluster states~\cite{zhang2006,menicucci2011,pfister2019} is motivated by the high scalability of this states with optical setups~\cite{yokoyama2013,chen2014,medeirosdearaujo2014,pfister2019,larsen2019,asavanant2019,wu2020}.
In this article we want to further analyze the relation between Gaussian cluster states%
~\footnote{Note that here, as in Ref.~\cite{menicucci2011}, the term ``cluster state'' indicates states with adjacency matrices corresponding to any weighted graph.} 
and multi-mode squeezed states~\cite{ma1990} (also called canonical graph states and H-graph states respectively~\cite{pfister2019}) which may have relevance for their practical implementation.

The possibility to generate multi-mode squeezed states with optical setups makes this class of Gaussian states very attractive as possible Gaussian cluster states for measurement-based universal quantum computation~\cite{menicucci2008}.
In fact, it as been shown~\cite{menicucci2007,pfister2019},
that any multi-mode squeezing transformation generates a cluster state. 
Here we show how to construct the multi-mode squeezing transformation which generates any given cluster state. We also specify the conditions under which a multi-mode squeezing transformation generates a cluster state. In particular, we describe in detail the mathematical relations between the adjacency matrix of the cluster and the matrix of squeezing interactions that constitute the multi-mode squeezing transformation.


A Gaussian cluster state $\ke{\Psi}$ is a zero eigenstate of the collective operators (the nullifiers)
\begin{eqnarray}\label{nullif}
x_j=-\ii\pt{b_j\,\ee^{\ii\,\theta_j}-b_j\da\,\ee^{-\ii\,\theta_j}}-\sum_{k=1}^N\,\AAA_{j,k}\pt{b_k\,\ee^{\ii\,\theta_k}+b_k\da\,\ee^{-\ii\,\theta_k}}\, ,
\end{eqnarray}
where, $b_j$ and $b_j\da$ are the annihilation and creation operators for $N$ bosonic modes, and $\AAA$ (real symmetric matrix) is the adjacency matrix which defines the cluster state, i.e. $x_j\,\ke{\Psi}=0$, $\forall\,j$. This means that these operators are infinitely squeezed. In practice it is possible to realize only approximated cluster states for which these operators are squeezed by a finite amount.
Note that in the definition of the nullifier we have included also any possible local rotation (phase shift) $\ee^{\ii\,\theta_j}$ which does not affect the global entanglement properties of the state.

In this work we show that every Gaussian cluster state defined by a real symmetric adjacency matrix $\AAA$ can be approximated by a multi-mode squeezed state $\ke{\Psi}$ defined in terms of a unitary transformation
\begin{eqnarray}\label{Uz}
U=\ee^{-\ii\,\frac{z}{2}\sum_{j,k=1}^N\pt{\ZZ_{j,k}b_j\da\,b_k\da+\ZZ_{j,k}\da\,b_j\,b_k}}
\end{eqnarray}
as
\begin{eqnarray}\label{Psiz}
\ke{\Psi}=U\ \ke{\bs 0}\ ,
\end{eqnarray}
where $z$ is a real positive number, $\ZZ$ is a complex, symmetric non-singular interaction matrix%
~\footnote{This definition of multi-mode squeezed states is more general then the one used in Refs.~\cite{menicucci2007,menicucci2011}, which corresponds to an imaginary matrix $\ZZ$. Moreover, note
that, considering the rotated nullifiers~\rp{nullif}, when, at the same time, we are considering a complex interaction matrix $\ZZ$, may seem redundant. In fact general local rotations can be equally well described by including the local phases $\ee^{\ii\,\theta_j}$ either in the definition of the nullifiers or in the complex interaction coefficients. However this notation is useful for a clearer discuss of the relation between cluster and multi-mode squeezed states when either one or the other is assumed fixed.}, and $\ke{\bs 0}$ is the vacuum.
Namely, we discuss the conditions under which the covariance matrix $\CC$ of the nullifiers over this state,
\begin{eqnarray}
\pg{\CC}_{j,k}=\br{\Psi}\frac{x_j\ x_k+x_k\ x_j}{2}\,\ke{\Psi}\ ,
\end{eqnarray}
approaches the null matrix in the limit of infinite $z$, i.e.
\begin{eqnarray}\label{limCz0}
\lim_{z\to\infty}\ \CC=0\ .
\end{eqnarray}
Notice that the modes described by the operators $b_j$ and $b_j\da$ are the physical modes that one
can control, manipulate and measure in a given experiment (such as the temporal or frequency modes discussed in Refs.~\cite{yokoyama2013,chen2014,medeirosdearaujo2014,pfister2019,larsen2019,asavanant2019}.
It is also important to point out that even if Eq~\rp{Uz} is not the most general Gaussian unitary transformation, since it does not include any $b_j\da\, b_k$ term, any zero-average Gaussian state can be generated from the vacuum by a multi-mode squeezing transformation of the form of Eq.~\rp{Uz}~\footnote{
Any general Gaussian unitary transformation can be decomposed as the product of a multi-mode squeezing transformation and a multi-mode interferometer~\cite{ma1990,cariolaro2016}. The latter, when applied to the vacuum, has no effect, and thus the state generated by a general Gaussian unitary transformation can be equivalently generated by a specific multi-mode squeezing transformation.}.

To be more specific, the central result of this work is the following theorem.

\noindent
\textit{Theorem:}
A state of the form~\rp{Uz}-\rp{Psiz}, with $\ZZ\,\in\,\mathbb{C}^{N\times N}$ non-singular and $\ZZ=\ZZ^T$, is a Gaussian cluster state with adjacency matrix $\AAA\,\in\,\mathbb{R}^{N\times N}$ with $\AAA=\AAA^T$, namely Eq.~\rp{limCz0} is true, if and only if the unitary matrix $\UU$, which enters into the polar decomposition of
$\ZZ$, defined by
\begin{eqnarray}\label{polardecomp}
\ZZ=\PP\,\UU
\end{eqnarray}
with $\PP=\pt{\ZZ\,\ZZ\da}^{1/2}$ hermitian positive definite (note that $\UU$ is symmetric because $\ZZ$ is symmetric), fulfills
\begin{eqnarray}\label{UN}
\UU=
-\ii\,\ee^{-\ii\,\Theta}\,\frac{\AAA-\ii\,\id}{\AAA+\ii\,\id}\,\ee^{-\ii\,\Theta}\ ,
\end{eqnarray}
where $\Theta$ is the diagonal matrix with entries $\Theta_{j,j}=\theta_j$.
In particular, in this case
\begin{eqnarray}\label{CC}
\CC&=&\pt{\AAA+\ii\,\id}\ \ee^{\ii\,\Theta}\,\ee^{-2\,z\,\PP}\,\ee^{-\ii\,\Theta}\ \pt{\AAA-\ii\,\id}
\\
&=&4\,\ee^{-\ii\,\Theta}\,\pt{\UU+\ii\,\ee^{-2\,\ii\,\Theta}}^{-1}\,\ee^{-2\,z\,\PP}\,\pt{\UU\da-\ii\,\ee^{2\,\ii\,\Theta}}^{-1}\,\ee^{\ii\,\Theta}\ .
\nn
\end{eqnarray}

\noindent{\it Proof:}
It is useful to express our equations in vector form in terms of the vector of mode operators $\vb=\pt{b_{1}\cdots,b_N,b_1\da\cdots,b_N\da}^T$ [with ${}^T$ indicating the transpose (which does not operates at the level of quantum operators) such that $\vb$ is a column vector]. Thereby,
the nullifiers~\rp{nullif} can be expressed, in vector form $\vx=\pt{x_1\cdots x_N}^T$, in terms of the $N\times 2\,N$ matrix
\begin{eqnarray}
\QQ=-\pt{\mat{cc}{\pq{\AAA+\ii\,\id}\,\ee^{\ii\,\Theta}\ \ &\ \ \pq{\AAA-\ii\,\id}\,\ee^{-\ii\,\Theta}}}
\end{eqnarray}
where $\id$ is the identity matrix, as
\begin{eqnarray}
\vx=\QQ\,\vb\ .
\end{eqnarray}
Correspondingly the covariance matrix is given by
\begin{eqnarray}
\CC&=&\br{\Psi}\,\QQ\,\frac{\vb\vb^T+\pt{\vb\vb^T}^T}{2}\,\QQ^T\,\ke{\Psi}\ .
\end{eqnarray}

The transformation of the mode operators under the effect of a Gaussian unitary operator $U$, which generates zero average Gaussian states [as Eq.~\rp{Uz}], can be expressed in terms of a $2\,N\times 2\,N$ Bogoliubov matrix $\BB$, such that
\begin{eqnarray}\label{BU}
U\da\ \vb_j U=\sum_{k=1}^{2\,N}\,\BB_{j,k}\,\vb_k
\end{eqnarray}
(where here $j$ is an index over the vector of operators and not an index of the modes).
Thereby we find
\begin{eqnarray}\label{CQBGBQ}
\CC&=&\QQ\,\BB\ \br{\bs 0}\,\frac{\vb\vb^T+\pt{\vb\vb^T}^T}{2}\,\ke{\bs 0}\ \BB^T\,\QQ^T
\nn\\
&=&\frac{1}{2}\QQ\ \BB\ \GG\ \BB^T\ \QQ^T\ ,
\end{eqnarray}
with $\GG=\matt{&\id}{\id&}$, where the missing blocks are null matrices.
In particular every Bogoliubov matrix $\BB$ takes the block structures
\begin{eqnarray}\label{BXY}
\BB=\matt{\XX&\YY}{\YY^*&\XX^*}\ ,
\end{eqnarray}
(note that with the symbol $\MM^*$ we indicate the matrix whose entries are the complex conjugates of the entries of $\MM$)
where the $N\times N$ complex matrices $\XX$ and $\YY$  fulfill the relations
\begin{eqnarray}\label{BogCond1}
\XX\,\XX\da-\YY\,\YY\da&=&\id
\nn\\
\XX\,\YY^T-\YY\,\XX^T&=&0\ ,
\end{eqnarray}
which are derived from the bosonic commutation relation.
By means of this, we find
\begin{eqnarray}
\QQ\,\BB=-\pt{\mat{cc}{\EE&\EE^*}}
\end{eqnarray}
with
\begin{eqnarray}\label{Az}
\EE&=&\pt{\AAA+\ii\id}\,\ee^{\ii\,\Theta}\,\XX+\pt{\AAA-\ii\id}\,\ee^{-\ii\,\Theta}\,\YY^*\ ,
\end{eqnarray}
so that
\begin{eqnarray}
\CC=\EE\ \EE\da\ ,
\end{eqnarray}
where the fact that $\CC$ is real can be shown using the relations~\rp{BogCond1}.
This expression for the covariance matrix of the nullifiers entails that
Eq.~\rp{limCz0} is equivalent to
\begin{eqnarray}\label{limAz0}
\lim_{z\to\infty}\ \EE=0\ .
\end{eqnarray}

In the case of the Gaussian transformation defined in Eq.~\rp{Uz}, the matrices $\XX$ and $\YY$ can be expressed in terms of the polar decomposition~\rp{polardecomp} as~\cite{ma1990}
\begin{eqnarray}\label{XY}
\XX&=&\cosh\pt{z\,\PP}
\nn\\
\YY&=&-\ii\sinh\pt{z\,\PP}\ \UU\ .
\end{eqnarray}
Thus, in general,
\begin{eqnarray}\label{limAz}
\lim_{z\to\infty}\,\EE&=&\frac{1}{2}\,\lim_{z\to\infty}\lpg{\pq{
\pt{\AAA+\ii\,\id}\,\ee^{\ii\,\Theta}+\ii\pt{\AAA-\ii\,\id}\,\ee^{-\ii\,\Theta}\,\UU^*}\,\ee^{z\,\PP}
}\nn\\&&\rpg{
+\pq{
\pt{\AAA+\ii\,\id}\,\ee^{\ii\,\Theta}-\ii\pt{\AAA-\ii\,\id}\,\ee^{-\ii\,\Theta}\,\UU^*}\,\ee^{-z\,\PP}
}\ ,
\end{eqnarray}
where we have used the fact that, since $\ZZ$ is symmetric, $\PP\,\UU=\UU\,\PP^*$.
So, finally, Eq.~\rp{limAz} is equal to the null matrix, namely Eqs.~\rp{limCz0} and \rp{limAz0} are true, if and only if $\UU$ is given by Eq.~\rp{UN}.
In this case we also find that  $\EE=\pt{\AAA+\ii\,\id}\,\ee^{\ii\,\Theta}\,\ee^{-z\,\PP}$
such that the covariance matrix is equal to Eq.~\rp{CC}.
$\hfill\blacksquare$

We also note that, given a symmetric unitary matrix $\UU$ of the form of Eq.~\rp{UN} with $\AAA$ real symmetric, and a hermitian matrix $\PP$,
then $\PP\,\UU$ is symmetric (i.e. $\PP\,\UU=\UU\,\PP^*$), if and only if
\begin{eqnarray}\label{condP}
\pt{\AAA+\ii\,\id}\,\ee^{\ii\,\Theta}\,\PP\,\ee^{-\ii\,\Theta}\,\pt{\AAA-\ii\,\id}\ \in \mathbb{R}^{N\times N}\ .
\end{eqnarray}
This proposition, in turn, entails that the covariance matrix in Eq.~\rp{CC} is real symmetric.

\paragraph*{The multi-mode squeezed state corresponding to a given cluster state.}
The result that we have demonstrated implies that, on the one hand, every cluster state with adjacency matrix $\AAA$ (and local mode rotations defined by the matrix $\Theta$) is approximated by the state generated by \rp{Uz} with $\ZZ$ given by the product of the unitary matrix in Eq.~\rp{UN} and any hermitian positive definite 
matrix $\PP$ that fulfill the relation~\rp{condP}.
In particular, given an adjacency matrix, there are infinite multi-mode squeezed states (corresponding to different $\PP$) that approximate the cluster state.
According to Ref.~\cite{gonzalez-arciniegas2020} a faithful approximation is one for which the covariance matrix of the nullifiers is diagonal. Our result shows that it is always possible to find such state.
Specifically, if we choose
\begin{eqnarray}\label{PPz}
\PP=\id+\ee^{-\ii\,\Theta}\,\frac{\ln\pt{\AAA^2+\id}}{2\,z}\,\ee^{\ii\,\Theta}\ ,
\end{eqnarray}
which is hermitian positive definite, and fulfill
the relations~\rp{condP} [when $\UU$ is given by Eq.~\rp{UN}], then we find $\ee^{-2\,z\,\PP}=\ee^{-2\,z}\,\ee^{-\ii\,\Theta}\,\pt{\AAA^2+\id}^{-1}\,\ee^{\ii\,\Theta}$ so that the covariance matrix of the nullifiers is
\begin{eqnarray}
\CC=\ee^{-2\,z}\,\id\ ,
\end{eqnarray}
for every real symmetric adjacency matrix $\AAA$.

\paragraph*{The cluster state corresponding to a given multi-mode squeezed state.}
On the other hand, given a general multi-mode squeezed state defined
by a complex symmetric matrix $\ZZ$, with polar decomposition $\ZZ=\PP\,\UU$, there exists a cluster state of which it is an approximation, when the relation~\rp{UN} can be solved for $\AAA$, namely when $\UU+\ii\,\ee^{-2\,\ii\,\Theta}$ is non-singular (and it is always possible to identify local rotations of angle $\theta_j$ for which this is true).
In this case, the multi-mode squeezed state is an approximation to the class of cluster states, corresponding to all possible local rotations which keep $\UU+\ii\,\ee^{-2\,\ii\,\Theta}$ non-singular, and defined by the adjacency matrix
\begin{eqnarray}\label{NU}
\AAA&=&-\ii\,\frac{\UU-\ii\,\ee^{-2\,\ii\,\Theta}}{\UU+\ii\,\ee^{-2\,\ii\,\Theta}}\ .
\end{eqnarray}
At fixed $\ZZ$, finding the cluster state [i.e. the adjacency matrix $\AAA$~\rp{NU}] which best approximates (that is the closest, by some measure to be determined, as a function of the local rotations $\Theta$, to) the multi-mode squeezed state is still an open question~\cite{menicucci2011,gonzalez-arciniegas2020}.
Note also that the unitary symmetric matrix $\ee^{\ii\,\Theta}\,\UU\,\ee^{\ii\,\Theta}$ can be expressed as $\ee^{\ii\,\Theta}\,\UU\,\ee^{\ii\,\Theta}=\ee^{\ii\,\KK}$, with $\KK$ real symmetric (because $\UU$ is symmetric), therefore we can express the adjacency matrix as $\AAA=-{\cos\pt{\KK}}/\pq{1+\sin\pt{\KK}}$,
which shows explicitly that $\AAA$ is real symmetric.

\paragraph*{The singular values of $\ZZ$ and the generation of a cluster state with many single-mode squeezed fields.}
We finally note that the structure of the cluster state is determined only by the unitary matrix of the polar decomposition~\rp{polardecomp}. The matrix $\PP$ determines instead how much the nullifiers~\rp{nullif} are actually squeezed.
To gain insight into the meaning of the matrix $\PP$
it is useful to analyze the transformation that generates the state~\rp{Uz} in terms of
the Bloch-Messiah reduction formula~\cite{braunstein2005,cariolaro2016a,cariolaro2016}, which shows how to generate any Gaussian state as a set of single mode squeezing operations followed by a set of beam splitter interactions and phase shifts~\cite{vanloock2007,gu2009,ferrini2015}.
In particular, the single mode-squeezing operations transform the mode operators according to $b_j\to\mu_j\,b_j+\nu_j\,b_j\da$, where $\mu_j$ and $\nu_j$ are the singular values of, respectively, the matrices $\XX$ and $\YY$ that constitute the Bogoliubov transformation corresponding to the state.
In our case the singular values of $\XX$ and $\YY$ in Eq.~\rp{XY} are given, in terms of the eigenvalues $\lambda_j$ of $\PP$ (that are the singular values of $\ZZ$), by $\mu_j=\cosh(z\,\lambda_j)$ and $\nu_j=\sinh(z\,\lambda_j)$ respectively.
So that, the eigenvalues of $\PP$ are a measure of the single-mode squeezing operations which enter in the Bloch-Messiah decomposition of the multi-mode squeezing transformation that generate the state. This means that a cluster state can be prepared by a set of squeezers with squeezing strength given by the eigenvalues of $\PP$ (i.e. the singular values of $\ZZ$), followed by a sequence of properly selected beam splitter interactions and phase shifts, determined according to the Bloch-Messiah reduction formula~\cite{vanloock2007,gu2009,ferrini2015,cariolaro2016a,cariolaro2016}.
We further highlight that with our approach we find the class of multi-mode squeezing transformations which generates a given cluster state. However the same state can be also generated by many generic Gaussian transformations as discussed in Ref.~\cite{vanloock2007,ferrini2015}, see appendix~\ref{app} for more details.

An interesting case is when $\PP=\id$, such that the state can be created starting by many equal single-mode squeezed states. In this case $\ZZ$ is equal to the unitary matrix $\ZZ=\UU$,
and the covariance matrix of the nullifiers is
\begin{eqnarray}
\CC&=&\pt{\AAA^2+\id}\ \ee^{-2\,z}\ .
\end{eqnarray}
Moreover, if we restrict to the case of
self-inverse adjacency matrix $\AAA^2=\id$, then the relations~\rp{limCz0} and \rp{limAz0} are true for
$\UU=-\ee^{-\ii\,\Theta}\,\AAA\,\ee^{-\ii\,\Theta}$. In this case the covariance matrix reduces to
\begin{eqnarray}\label{CCb}
\CC&=&2\,\ee^{-2\,z}\ \id\ .
\end{eqnarray}
We note that these Gaussian cluster states, with self-inverse adjacency matrix, are the states discussed and generated in~\cite{menicucci2007,menicucci2008,zaidi2008,flammia2009,menicucci2011a,menicucci2011,yokoyama2013,
chen2014,wang2014a,alexander2016b,larsen2019}.


\paragraph*{Conclusions.}
We have discussed the connections between multi-mode squeezed states and cluster states showing that any cluster state
(not only bipartite ones~\cite{menicucci2007})
can be approximated by a multi-mode squeezed state. Namely, given any cluster state, we have shown how to determine a multi-mode squeezed state for which the covariance matrix of the nullifiers (that determine the cluster) approaches the null matrix in the limit of infinite squeezing.
The choice of the multi-mode squeezed state corresponding to a cluster is not unique, and
we have also shown that for any cluster state, the multi-mode squeezed state can always be chosen such that the corresponding covariance matrix of the nullifiers is diagonal.

These findings may help to identify additional interferometric strategies  to entangle either temporal or frequency modes alternative to those realized in~\cite{yokoyama2013,chen2014,medeirosdearaujo2014,larsen2019,asavanant2019} and that could achieve more general cluster geometries.

\paragraph*{Acknowledgments}
We acknowledge the support of the European Union Horizon 2020 Programme for Research and Innovation through the Project No. 732894 (FET Proactive HOT), and the Project No. 862644 (FET Open QUARTET).

\appendix

\section{Relation between our result and Refs.~\cite{vanloock2007,ferrini2015}}
\label{app}

The Bloch-Messiah formula~\cite{braunstein2005,cariolaro2016a,cariolaro2016} shows that it is
possible to generate any Gaussian state by manipulating many squeezed mode with a multiport interferometer~\cite{braunstein2005,vanloock2007,gu2009,cariolaro2016a,cariolaro2016}. 
In detail, the Bloch-Messiah formula shows that it is always possible to decompose the matrices $\XX$ and $\YY$ that constitute a general Bogoliubov matrix $\BB$~\rp{BXY} in terms of the single value decompositions 
\begin{eqnarray}\label{BMVW}
\XX&=&\VV\ \DD_x\  \WW\da
\nn\\
\YY&=&\VV\ \DD_y\  \WW^T
\end{eqnarray}
where $\VV$ and $\WW$ are two unitary matrices, and $\DD_x$ and $\DD_y$ diagonal. In particular $\DD_x$ and $\DD_y$ can be expressed in terms of a diagonal matrix $z\,\DD$ as $\DD_x=\cosh(z\DD)$ and $\DD_y=\sinh(z\DD)$. Note that this decomposition is general 
[it is valid for any Bogoliubov matrix $\BB$, which corresponds to any general Gaussian unitary transformation $U_{\BB}=\ee^{-\ii\sum_{j,k=1}^N\pt{\ZZ_{j,k}\ b_j\da b_k\da+\ZZ_{j,k}^*\ b_j b_k+\SSS_{j,k}\ b_j\da b_k+\SSS_{j,k}^*\ b_j b_k\da}}$, with $\ZZ=\ZZ^T$ and $\SSS=\SSS\da$, according to the relation 
%
$\BB\,\vb=U_\BB\ \vb\ U_\BB\da=\pt{
U_\BB\ b_1\ U_\BB\da,\cdots U_\BB\ b_N\ U_\BB\da,
U_\BB\ b_1\da\ U_\BB\da,\cdots U_\BB\ b_N\da\ U_\BB\da
}^T$
], and it is different from Eq.~\rp{XY} which is valid only for a multi-mode squeezing transformation~\rp{Uz}. Thus, a general Bogoliubov matrix~\rp{BXY}  can be decomposed as the product of three matrices $\BB=\BB_\VV\ \BB_\DD\ \BB_\WW$, 
\begin{eqnarray}
\BB_\VV=\pt{\mat{cc}{\VV&\\&\VV^*}},\  
\BB_\DD=\pt{\mat{cc}{\DD_x&\DD_y\\\DD_y&\DD_x}},\ 
\BB_\WW=\pt{\mat{cc}{\WW\da&\\&\WW^T}},
\nn\\
\end{eqnarray}
where the missing blocks are null matrices, and $\BB_\DD$ describes the squeezing of all the modes, while $\BB_\VV$ and $\BB_\WW$ represent multi-port interferometers. Moreover, since each Bogoliubov matrix is related to a Gaussian unitary transformation $U_{\BB_x}$ (with $x\in\pg{\VV,\DD,\WW}$)
by the relation $\BB_x\,\vb=U_{\BB_x}\ \vb\ U_{\BB_x}\da$, one finds $U_\BB=U_{\BB_\VV}\,U_{\BB_\DD}\,U_{\BB_\WW}$, so that the state generated from the vacuum $\ke{0}$ by a general Gaussian unitary transformation can be always written as $U_\BB\,\ke{0}=U_{\BB_\VV}\,U_{\BB_\DD}\,\ke{0}$, where $U_{\BB_\WW}$ has no effect on the vacuum~\cite{vanloock2007,gu2009}.

Exploiting this result it is possible to identify a Gaussian unitary transformation which generates any cluster state, see Refs.~\cite{vanloock2007,ferrini2015}.
Let us now rephrase this result following our notation.
Using the decomposition~\rp{BMVW} in Eq.~\rp{Az}  we find 
\begin{eqnarray}
\lim_{z\to\infty}\,\EE&=&\frac{1}{2}\,\lim_{z\to\infty}\lpg{\pq{
\pt{\AAA+\ii\,\id}\,\ee^{\ii\,\Theta}\,\VV+\ii\pt{\AAA-\ii\,\id}\,\ee^{-\ii\,\Theta}\,\VV^*}\,\ee^{z\DD}
}\nn\\&&\rpg{
+\pq{
\pt{\AAA+\ii\,\id}\,\ee^{\ii\,\Theta}\,\VV-\ii\pt{\AAA-\ii\,\id}\,\ee^{-\ii\,\Theta}\,\VV^*}\,\ee^{-z\DD}
}\,\WW\da\ ,
\nn\\
\end{eqnarray} 
which is zero (meaning that the corresponding state is a Gaussian cluster state) if and only if 
\begin{eqnarray}
\pt{\AAA+\ii\,\id}\,\ee^{\ii\,\Theta}\,\VV+\ii\pt{\AAA-\ii\,\id}\,\ee^{-\ii\,\Theta}\,\VV^*=0 ,
\end{eqnarray}
and, introducing the real and imaginary parts of $\ee^{\ii\,\Theta}\,\VV$ such that $\ee^{\ii\,\Theta}\,\VV=\VV_r+\ii\,\VV_i$, we find
that this expression can be rewritten as
\begin{eqnarray}
\VV_i-\AAA\,\VV_r=0
\end{eqnarray}
which is equivalent (apart form the detail of the local rotations $\ee^{\ii\,\Theta}$) to the Eq.~(17) of Ref.~\cite{vanloock2007} and to the Eq.~(5) of Ref.~\cite{ferrini2015}. 
In particular this entails that $\VV_i=\AAA\,\VV_r$, so that $\ee^{\ii\,\Theta}\,\VV=\pt{\id-\ii\,\AAA}\,\VV_r$, and since $\ee^{\ii\,\Theta}\,\VV$ is unitary ($\ee^{\ii\,\Theta}\,\VV\,\VV\da\,\ee^{-\ii\,\Theta}=\id$) we find $\VV_r\,\VV_r^T=\pt{\id+\AAA^2}^{-1}$, which is equivalent to Eq.~(9) of Ref.~\cite{ferrini2015}. Correspondingly, we find $\VV_r=\pt{\id+\AAA^2}^{-1/2}\,\OO$, where $\OO$ is a generic real orthogonal matrix. Thus, we find	
\begin{eqnarray}\label{VVFerrini}
\VV=\ee^{-\ii\,\Theta}\,\pt{\id+\ii\,\AAA}\ \pt{\AAA^2+\id}^{-1/2}\,\OO\ ,
\end{eqnarray}
which is equivalent to the Eq.~(12) of Ref.~\cite{ferrini2015}. 
This equation shows that any Gaussian state generated by a unitary transformation, whose corresponding Bogoliubov matrix is decomposed as in Eq.~\rp{BMVW}, with the unitary matrix $\VV$ which fulfills Eq.~\rp{VVFerrini}, is a faithful approximation of a Gaussian cluster state with adjacency matrix $\AAA$.

Let us now analyze how this known result is related to the result that we discuss in this work.
In order to compare our results with Eq.~\rp{VVFerrini}, we have to find a relation between the Bloch-Messiah decomposition~\rp{BMVW}  and the decomposition~\rp{XY} (which is valid only for multi-mode squeezed states).
A singular value decomposition of the matrices $\XX$ and $\YY$ in Eq.~\rp{XY} is obtained by exploiting the diagonalization of the matrix $\PP$, that is 
$\PP=\TT\da\ \DD_\PP\ \TT$ (with $\TT$ unitary and $\DD_\PP$ diagonal), and which gives 
\begin{eqnarray}\label{XYTDTU}
\XX&=&\TT\da\ \cosh(\DD_\PP)\ \TT
\nn\\
\YY&=&\TT\da\ \sinh(\DD_\PP)\ \TT \pt{-\ii\,\UU}\ .
\end{eqnarray}
This is not in the form of Eq.\rp{BMVW}. In fact the singular value decomposition of a matrix is not unique, and Eq.\rp{BMVW} is made by the pair of singular value decompositions which are written in terms of only two unitary matrices $\VV$ and $\WW$. Eq.~\rp{XYTDTU} can be cast in this peculiar form following the procedure discussed in Refs.~\cite{cariolaro2016a,cariolaro2016}.
Specifically, one can find a balancing matrix $\RR$ (unitary) 
such that Eq.\rp{XYTDTU} can be cast in the form of Eq.~\rp{BMVW} with 
\begin{eqnarray}\label{VV}
\VV&=&\TT\da\,\RR
\\
\WW&=&-\ii\,\UU\,\TT^T\,\RR^*
\\
z\,\DD&=&\DD_\PP\ .
\end{eqnarray}
In particular, the balancing matrix is the one given by the Autonne–Takagi factorization~\cite{cariolaro2016a,cariolaro2016} of the symmetric unitary $-\ii \TT\ \UU\ \TT^T$, such that
\begin{eqnarray}
-\ii \TT\ \UU\ \TT^T= \RR\ \RR^T\ .
\end{eqnarray}
Equivalently, the unitary $\TT\da\ \RR$ is given by the Autonne–Takagi factorization of $-\ii\,\UU$, that is 
\begin{eqnarray}\label{UTR}
-\ii \UU= \TT\da\,\RR\ \RR^T\,\TT^*\ .
\end{eqnarray}
Thereby, using Eqs.~\rp{VV} and \rp{UTR} we find 
\begin{eqnarray}\label{UiVV}
\UU=\ii\,\VV\ \VV^T \ .
\end{eqnarray}   
This relation indicates that given a general Gaussian unitary transformation $U_\BB$, with corresponding Bogoliubov matrix $\BB$ which can be decomposed in terms of the matrices $\VV$ and $\DD$ as in Eq.~\rp{BMVW}, 
generates, from the vacuum, the same state generated by a multi-mode squeezing transformations $U$~\rp{Uz}, characterized by an interaction matrix $\ZZ$ with polar decomposition~\rp{polardecomp} expressed in terms of the unitary matrix $\UU$ given in Eq.~\rp{UiVV} and a matrix $\PP$ with eigenvalues equal to the diagonal elements of the diagonal matrix $\DD$.

In particular, if we consider a general Gaussian unitary transformation which generates an approximation of a cluster state with adjacency matrix $\AAA$, and which is characterized by Eq.~\rp{BMVW}, with $\VV$ given in Eq.~\rp{VVFerrini}, we find that the same approximation is generated by a multi-mode squeezing transformation characterized by a matrix $\UU$~\rp{UiVV} given by
\begin{eqnarray}
\UU&=&\ii\,\ee^{-\ii\,\Theta}\,\pt{\id+\ii\,\AAA}\ \pt{\AAA^2+\id}^{-1}\,\pt{\id+\ii\,\AAA} \,\ee^{-\ii\,\Theta}
\end{eqnarray}   
which is equal to Eq.~\rp{UN}. This is an alternative derivation, which makes use of the Bloch-Messiah decomposition, of the theorem discussed in the main text.


\begin{thebibliography}{29}%
\makeatletter
\providecommand \@ifxundefined [1]{%
 \@ifx{#1\undefined}
}%
\providecommand \@ifnum [1]{%
 \ifnum #1\expandafter \@firstoftwo
 \else \expandafter \@secondoftwo
 \fi
}%
\providecommand \@ifx [1]{%
 \ifx #1\expandafter \@firstoftwo
 \else \expandafter \@secondoftwo
 \fi
}%
\providecommand \natexlab [1]{#1}%
\providecommand \enquote  [1]{``#1''}%
\providecommand \bibnamefont  [1]{#1}%
\providecommand \bibfnamefont [1]{#1}%
\providecommand \citenamefont [1]{#1}%
\providecommand \href@noop [0]{\@secondoftwo}%
\providecommand \href [0]{\begingroup \@sanitize@url \@href}%
\providecommand \@href[1]{\@@startlink{#1}\@@href}%
\providecommand \@@href[1]{\endgroup#1\@@endlink}%
\providecommand \@sanitize@url [0]{\catcode `\\12\catcode `\$12\catcode
  `\&12\catcode `\#12\catcode `\^12\catcode `\_12\catcode `\%12\relax}%
\providecommand \@@startlink[1]{}%
\providecommand \@@endlink[0]{}%
\providecommand \url  [0]{\begingroup\@sanitize@url \@url }%
\providecommand \@url [1]{\endgroup\@href {#1}{\urlprefix }}%
\providecommand \urlprefix  [0]{URL }%
\providecommand \Eprint [0]{\href }%
\providecommand \doibase [0]{http://dx.doi.org/}%
\providecommand \selectlanguage [0]{\@gobble}%
\providecommand \bibinfo  [0]{\@secondoftwo}%
\providecommand \bibfield  [0]{\@secondoftwo}%
\providecommand \translation [1]{[#1]}%
\providecommand \BibitemOpen [0]{}%
\providecommand \bibitemStop [0]{}%
\providecommand \bibitemNoStop [0]{.\EOS\space}%
\providecommand \EOS [0]{\spacefactor3000\relax}%
\providecommand \BibitemShut  [1]{\csname bibitem#1\endcsname}%
\let\auto@bib@innerbib\@empty
\bibitem [{\citenamefont {Raussendorf}\ and\ \citenamefont
  {Briegel}(2001)}]{raussendorf2001}%
  \BibitemOpen
  \bibfield  {author} {\bibinfo {author} {\bibfnamefont {Robert}\ \bibnamefont
  {Raussendorf}}\ and\ \bibinfo {author} {\bibfnamefont {Hans~J.}\ \bibnamefont
  {Briegel}},\ }\bibfield  {title} {\enquote {\bibinfo {title} {A {{One}}-{{Way
  Quantum Computer}}},}\ }\href {\doibase 10.1103/PhysRevLett.86.5188}
  {\bibfield  {journal} {\bibinfo  {journal} {Phys. Rev. Lett.}\ }\textbf
  {\bibinfo {volume} {86}},\ \bibinfo {pages} {5188--5191} (\bibinfo {year}
  {2001})}\BibitemShut {NoStop}%
\bibitem [{\citenamefont {Menicucci}\ \emph {et~al.}(2006)\citenamefont
  {Menicucci}, \citenamefont {{van Loock}}, \citenamefont {Gu}, \citenamefont
  {Weedbrook}, \citenamefont {Ralph},\ and\ \citenamefont
  {Nielsen}}]{menicucci2006}%
  \BibitemOpen
  \bibfield  {author} {\bibinfo {author} {\bibfnamefont {Nicolas~C.}\
  \bibnamefont {Menicucci}}, \bibinfo {author} {\bibfnamefont {Peter}\
  \bibnamefont {{van Loock}}}, \bibinfo {author} {\bibfnamefont {Mile}\
  \bibnamefont {Gu}}, \bibinfo {author} {\bibfnamefont {Christian}\
  \bibnamefont {Weedbrook}}, \bibinfo {author} {\bibfnamefont {Timothy~C.}\
  \bibnamefont {Ralph}}, \ and\ \bibinfo {author} {\bibfnamefont {Michael~A.}\
  \bibnamefont {Nielsen}},\ }\bibfield  {title} {\enquote {\bibinfo {title}
  {Universal {{Quantum Computation}} with {{Continuous}}-{{Variable Cluster
  States}}},}\ }\href {\doibase 10.1103/PhysRevLett.97.110501} {\bibfield
  {journal} {\bibinfo  {journal} {Phys. Rev. Lett.}\ }\textbf {\bibinfo
  {volume} {97}},\ \bibinfo {pages} {110501} (\bibinfo {year}
  {2006})}\BibitemShut {NoStop}%
\bibitem [{\citenamefont {Zhang}\ and\ \citenamefont
  {Braunstein}(2006)}]{zhang2006}%
  \BibitemOpen
  \bibfield  {author} {\bibinfo {author} {\bibfnamefont {Jing}\ \bibnamefont
  {Zhang}}\ and\ \bibinfo {author} {\bibfnamefont {Samuel~L.}\ \bibnamefont
  {Braunstein}},\ }\bibfield  {title} {\enquote {\bibinfo {title}
  {Continuous-variable {{Gaussian}} analog of cluster states},}\ }\href
  {\doibase 10.1103/PhysRevA.73.032318} {\bibfield  {journal} {\bibinfo
  {journal} {Phys. Rev. A}\ }\textbf {\bibinfo {volume} {73}},\ \bibinfo
  {pages} {032318} (\bibinfo {year} {2006})}\BibitemShut {NoStop}%
\bibitem [{\citenamefont {Menicucci}\ \emph {et~al.}(2011)\citenamefont
  {Menicucci}, \citenamefont {Flammia},\ and\ \citenamefont {{van
  Loock}}}]{menicucci2011}%
  \BibitemOpen
  \bibfield  {author} {\bibinfo {author} {\bibfnamefont {Nicolas~C.}\
  \bibnamefont {Menicucci}}, \bibinfo {author} {\bibfnamefont {Steven~T.}\
  \bibnamefont {Flammia}}, \ and\ \bibinfo {author} {\bibfnamefont {Peter}\
  \bibnamefont {{van Loock}}},\ }\bibfield  {title} {\enquote {\bibinfo {title}
  {Graphical calculus for {{Gaussian}} pure states},}\ }\href {\doibase
  10.1103/PhysRevA.83.042335} {\bibfield  {journal} {\bibinfo  {journal} {Phys.
  Rev. A}\ }\textbf {\bibinfo {volume} {83}},\ \bibinfo {pages} {042335}
  (\bibinfo {year} {2011})}\BibitemShut {NoStop}%
\bibitem [{\citenamefont {Pfister}(2019)}]{pfister2019}%
  \BibitemOpen
  \bibfield  {author} {\bibinfo {author} {\bibfnamefont {Olivier}\ \bibnamefont
  {Pfister}},\ }\bibfield  {title} {\enquote {\bibinfo {title}
  {Continuous-variable quantum computing in the quantum optical frequency
  comb},}\ }\href {\doibase 10.1088/1361-6455/ab526f} {\bibfield  {journal}
  {\bibinfo  {journal} {J. Phys. B: At. Mol. Opt. Phys.}\ }\textbf {\bibinfo
  {volume} {53}},\ \bibinfo {pages} {012001} (\bibinfo {year}
  {2019})}\BibitemShut {NoStop}%
\bibitem [{\citenamefont {Yokoyama}\ \emph {et~al.}(2013)\citenamefont
  {Yokoyama}, \citenamefont {Ukai}, \citenamefont {Armstrong}, \citenamefont
  {Sornphiphatphong}, \citenamefont {Kaji}, \citenamefont {Suzuki},
  \citenamefont {Yoshikawa}, \citenamefont {Yonezawa}, \citenamefont
  {Menicucci},\ and\ \citenamefont {Furusawa}}]{yokoyama2013}%
  \BibitemOpen
  \bibfield  {author} {\bibinfo {author} {\bibfnamefont {Shota}\ \bibnamefont
  {Yokoyama}}, \bibinfo {author} {\bibfnamefont {Ryuji}\ \bibnamefont {Ukai}},
  \bibinfo {author} {\bibfnamefont {Seiji~C.}\ \bibnamefont {Armstrong}},
  \bibinfo {author} {\bibfnamefont {Chanond}\ \bibnamefont {Sornphiphatphong}},
  \bibinfo {author} {\bibfnamefont {Toshiyuki}\ \bibnamefont {Kaji}}, \bibinfo
  {author} {\bibfnamefont {Shigenari}\ \bibnamefont {Suzuki}}, \bibinfo
  {author} {\bibfnamefont {Jun-ichi}\ \bibnamefont {Yoshikawa}}, \bibinfo
  {author} {\bibfnamefont {Hidehiro}\ \bibnamefont {Yonezawa}}, \bibinfo
  {author} {\bibfnamefont {Nicolas~C.}\ \bibnamefont {Menicucci}}, \ and\
  \bibinfo {author} {\bibfnamefont {Akira}\ \bibnamefont {Furusawa}},\
  }\bibfield  {title} {\enquote {\bibinfo {title} {Ultra-large-scale
  continuous-variable cluster states multiplexed in the time domain},}\ }\href
  {\doibase 10.1038/nphoton.2013.287} {\bibfield  {journal} {\bibinfo
  {journal} {Nat Photon}\ }\textbf {\bibinfo {volume} {7}},\ \bibinfo {pages}
  {982--986} (\bibinfo {year} {2013})}\BibitemShut {NoStop}%
\bibitem [{\citenamefont {Chen}\ \emph {et~al.}(2014)\citenamefont {Chen},
  \citenamefont {Menicucci},\ and\ \citenamefont {Pfister}}]{chen2014}%
  \BibitemOpen
  \bibfield  {author} {\bibinfo {author} {\bibfnamefont {Moran}\ \bibnamefont
  {Chen}}, \bibinfo {author} {\bibfnamefont {Nicolas~C.}\ \bibnamefont
  {Menicucci}}, \ and\ \bibinfo {author} {\bibfnamefont {Olivier}\ \bibnamefont
  {Pfister}},\ }\bibfield  {title} {\enquote {\bibinfo {title} {Experimental
  {{Realization}} of {{Multipartite Entanglement}} of 60 {{Modes}} of a
  {{Quantum Optical Frequency Comb}}},}\ }\href {\doibase
  10.1103/PhysRevLett.112.120505} {\bibfield  {journal} {\bibinfo  {journal}
  {Phys. Rev. Lett.}\ }\textbf {\bibinfo {volume} {112}},\ \bibinfo {pages}
  {120505} (\bibinfo {year} {2014})}\BibitemShut {NoStop}%
\bibitem [{\citenamefont {{Medeiros de Ara{\'u}jo}}\ \emph
  {et~al.}(2014)\citenamefont {{Medeiros de Ara{\'u}jo}}, \citenamefont
  {Roslund}, \citenamefont {Cai}, \citenamefont {Ferrini}, \citenamefont
  {Fabre},\ and\ \citenamefont {Treps}}]{medeirosdearaujo2014}%
  \BibitemOpen
  \bibfield  {author} {\bibinfo {author} {\bibfnamefont {R.}~\bibnamefont
  {{Medeiros de Ara{\'u}jo}}}, \bibinfo {author} {\bibfnamefont
  {J.}~\bibnamefont {Roslund}}, \bibinfo {author} {\bibfnamefont
  {Y.}~\bibnamefont {Cai}}, \bibinfo {author} {\bibfnamefont {G.}~\bibnamefont
  {Ferrini}}, \bibinfo {author} {\bibfnamefont {C.}~\bibnamefont {Fabre}}, \
  and\ \bibinfo {author} {\bibfnamefont {N.}~\bibnamefont {Treps}},\ }\bibfield
   {title} {\enquote {\bibinfo {title} {Full characterization of a highly
  multimode entangled state embedded in an optical frequency comb using pulse
  shaping},}\ }\href {\doibase 10.1103/PhysRevA.89.053828} {\bibfield
  {journal} {\bibinfo  {journal} {Phys. Rev. A}\ }\textbf {\bibinfo {volume}
  {89}},\ \bibinfo {pages} {053828} (\bibinfo {year} {2014})}\BibitemShut
  {NoStop}%
\bibitem [{\citenamefont {Larsen}\ \emph {et~al.}(2019)\citenamefont {Larsen},
  \citenamefont {Guo}, \citenamefont {Breum}, \citenamefont
  {{Neergaard-Nielsen}},\ and\ \citenamefont {Andersen}}]{larsen2019}%
  \BibitemOpen
  \bibfield  {author} {\bibinfo {author} {\bibfnamefont {Mikkel~V.}\
  \bibnamefont {Larsen}}, \bibinfo {author} {\bibfnamefont {Xueshi}\
  \bibnamefont {Guo}}, \bibinfo {author} {\bibfnamefont {Casper~R.}\
  \bibnamefont {Breum}}, \bibinfo {author} {\bibfnamefont {Jonas~S.}\
  \bibnamefont {{Neergaard-Nielsen}}}, \ and\ \bibinfo {author} {\bibfnamefont
  {Ulrik~L.}\ \bibnamefont {Andersen}},\ }\bibfield  {title} {\enquote
  {\bibinfo {title} {Deterministic generation of a two-dimensional cluster
  state},}\ }\href {\doibase 10.1126/science.aay4354} {\bibfield  {journal}
  {\bibinfo  {journal} {Science}\ }\textbf {\bibinfo {volume} {366}},\ \bibinfo
  {pages} {369--372} (\bibinfo {year} {2019})}\BibitemShut {NoStop}%
\bibitem [{\citenamefont {Asavanant}\ \emph {et~al.}(2019)\citenamefont
  {Asavanant}, \citenamefont {Shiozawa}, \citenamefont {Yokoyama},
  \citenamefont {Charoensombutamon}, \citenamefont {Emura}, \citenamefont
  {Alexander}, \citenamefont {Takeda}, \citenamefont {Yoshikawa}, \citenamefont
  {Menicucci}, \citenamefont {Yonezawa},\ and\ \citenamefont
  {Furusawa}}]{asavanant2019}%
  \BibitemOpen
  \bibfield  {author} {\bibinfo {author} {\bibfnamefont {Warit}\ \bibnamefont
  {Asavanant}}, \bibinfo {author} {\bibfnamefont {Yu}~\bibnamefont {Shiozawa}},
  \bibinfo {author} {\bibfnamefont {Shota}\ \bibnamefont {Yokoyama}}, \bibinfo
  {author} {\bibfnamefont {Baramee}\ \bibnamefont {Charoensombutamon}},
  \bibinfo {author} {\bibfnamefont {Hiroki}\ \bibnamefont {Emura}}, \bibinfo
  {author} {\bibfnamefont {Rafael~N.}\ \bibnamefont {Alexander}}, \bibinfo
  {author} {\bibfnamefont {Shuntaro}\ \bibnamefont {Takeda}}, \bibinfo {author}
  {\bibfnamefont {Jun-ichi}\ \bibnamefont {Yoshikawa}}, \bibinfo {author}
  {\bibfnamefont {Nicolas~C.}\ \bibnamefont {Menicucci}}, \bibinfo {author}
  {\bibfnamefont {Hidehiro}\ \bibnamefont {Yonezawa}}, \ and\ \bibinfo {author}
  {\bibfnamefont {Akira}\ \bibnamefont {Furusawa}},\ }\bibfield  {title}
  {\enquote {\bibinfo {title} {Generation of time-domain-multiplexed
  two-dimensional cluster state},}\ }\href {\doibase 10.1126/science.aay2645}
  {\bibfield  {journal} {\bibinfo  {journal} {Science}\ }\textbf {\bibinfo
  {volume} {366}},\ \bibinfo {pages} {373--376} (\bibinfo {year}
  {2019})}\BibitemShut {NoStop}%
\bibitem [{\citenamefont {Wu}\ \emph {et~al.}(2020)\citenamefont {Wu},
  \citenamefont {Alexander}, \citenamefont {Liu},\ and\ \citenamefont
  {Zhang}}]{wu2020}%
  \BibitemOpen
  \bibfield  {author} {\bibinfo {author} {\bibfnamefont {Bo-Han}\ \bibnamefont
  {Wu}}, \bibinfo {author} {\bibfnamefont {Rafael~N.}\ \bibnamefont
  {Alexander}}, \bibinfo {author} {\bibfnamefont {Shuai}\ \bibnamefont {Liu}},
  \ and\ \bibinfo {author} {\bibfnamefont {Zheshen}\ \bibnamefont {Zhang}},\
  }\bibfield  {title} {\enquote {\bibinfo {title} {Quantum computing with
  multidimensional continuous-variable cluster states in a scalable photonic
  platform},}\ }\href {\doibase 10.1103/PhysRevResearch.2.023138} {\bibfield
  {journal} {\bibinfo  {journal} {Phys. Rev. Research}\ }\textbf {\bibinfo
  {volume} {2}},\ \bibinfo {pages} {023138} (\bibinfo {year}
  {2020})}\BibitemShut {NoStop}%
\bibitem [{Note1()}]{Note1}%
  \BibitemOpen
  \bibinfo {note} {Note that here, as in Ref.~\cite {menicucci2011}, the term
  ``cluster state'' indicates states with adjacency matrices corresponding to
  any weighted graph.}\BibitemShut {Stop}%
\bibitem [{\citenamefont {Ma}\ and\ \citenamefont {Rhodes}(1990)}]{ma1990}%
  \BibitemOpen
  \bibfield  {author} {\bibinfo {author} {\bibfnamefont {Xin}\ \bibnamefont
  {Ma}}\ and\ \bibinfo {author} {\bibfnamefont {William}\ \bibnamefont
  {Rhodes}},\ }\bibfield  {title} {\enquote {\bibinfo {title} {Multimode
  squeeze operators and squeezed states},}\ }\href {\doibase
  10.1103/PhysRevA.41.4625} {\bibfield  {journal} {\bibinfo  {journal} {Phys.
  Rev. A}\ }\textbf {\bibinfo {volume} {41}},\ \bibinfo {pages} {4625--4631}
  (\bibinfo {year} {1990})}\BibitemShut {NoStop}%
\bibitem [{\citenamefont {Menicucci}\ \emph {et~al.}(2008)\citenamefont
  {Menicucci}, \citenamefont {Flammia},\ and\ \citenamefont
  {Pfister}}]{menicucci2008}%
  \BibitemOpen
  \bibfield  {author} {\bibinfo {author} {\bibfnamefont {Nicolas~C.}\
  \bibnamefont {Menicucci}}, \bibinfo {author} {\bibfnamefont {Steven~T.}\
  \bibnamefont {Flammia}}, \ and\ \bibinfo {author} {\bibfnamefont {Olivier}\
  \bibnamefont {Pfister}},\ }\bibfield  {title} {\enquote {\bibinfo {title}
  {One-{{Way Quantum Computing}} in the {{Optical Frequency Comb}}},}\ }\href
  {\doibase 10.1103/PhysRevLett.101.130501} {\bibfield  {journal} {\bibinfo
  {journal} {Phys. Rev. Lett.}\ }\textbf {\bibinfo {volume} {101}},\ \bibinfo
  {pages} {130501} (\bibinfo {year} {2008})}\BibitemShut {NoStop}%
\bibitem [{\citenamefont {Menicucci}\ \emph {et~al.}(2007)\citenamefont
  {Menicucci}, \citenamefont {Flammia}, \citenamefont {Zaidi},\ and\
  \citenamefont {Pfister}}]{menicucci2007}%
  \BibitemOpen
  \bibfield  {author} {\bibinfo {author} {\bibfnamefont {Nicolas~C.}\
  \bibnamefont {Menicucci}}, \bibinfo {author} {\bibfnamefont {Steven~T.}\
  \bibnamefont {Flammia}}, \bibinfo {author} {\bibfnamefont {Hussain}\
  \bibnamefont {Zaidi}}, \ and\ \bibinfo {author} {\bibfnamefont {Olivier}\
  \bibnamefont {Pfister}},\ }\bibfield  {title} {\enquote {\bibinfo {title}
  {Ultracompact generation of continuous-variable cluster states},}\ }\href
  {\doibase 10.1103/PhysRevA.76.010302} {\bibfield  {journal} {\bibinfo
  {journal} {Phys. Rev. A}\ }\textbf {\bibinfo {volume} {76}},\ \bibinfo
  {pages} {010302(R)} (\bibinfo {year} {2007})}\BibitemShut {NoStop}%
\bibitem [{Note2()}]{Note2}%
  \BibitemOpen
  \bibinfo {note} {This definition of multi-mode squeezed states is more
  general then the one used in Refs.~\cite {menicucci2007,menicucci2011}, which
  corresponds to an imaginary matrix ${\protect \cal Z}$. Moreover, note that,
  considering the rotated nullifiers~(\ref {nullif}), when, at the same time,
  we are considering a complex interaction matrix ${\protect \cal Z}$, may seem
  redundant. In fact general local rotations can be equally well described by
  including the local phases ${\protect \rm e}^{{\protect \rm i}\protect
  \tmspace +\thinmuskip {.1667em}\theta _j}$ either in the definition of the
  nullifiers or in the complex interaction coefficients. However this notation
  is useful for a clearer discuss of the relation between cluster and
  multi-mode squeezed states when either one or the other is assumed
  fixed.}\BibitemShut {Stop}%
\bibitem [{Note3()}]{Note3}%
  \BibitemOpen
  \bibinfo {note} {Any general Gaussian unitary transformation can be
  decomposed as the product of a multi-mode squeezing transformation and a
  multi-mode interferometer~\cite {ma1990,cariolaro2016}. The latter, when
  applied to the vacuum, has no effect, and thus the state generated by a
  general Gaussian unitary transformation can be equivalently generated by a
  specific multi-mode squeezing transformation.}\BibitemShut {Stop}%
\bibitem [{\citenamefont {{Gonz{\'a}lez-Arciniegas}}\ \emph
  {et~al.}(2020)\citenamefont {{Gonz{\'a}lez-Arciniegas}}, \citenamefont
  {Nussenzveig}, \citenamefont {Martinelli},\ and\ \citenamefont
  {Pfister}}]{gonzalez-arciniegas2020}%
  \BibitemOpen
  \bibfield  {author} {\bibinfo {author} {\bibfnamefont {Carlos}\ \bibnamefont
  {{Gonz{\'a}lez-Arciniegas}}}, \bibinfo {author} {\bibfnamefont {Paulo}\
  \bibnamefont {Nussenzveig}}, \bibinfo {author} {\bibfnamefont {Marcelo}\
  \bibnamefont {Martinelli}}, \ and\ \bibinfo {author} {\bibfnamefont
  {Olivier}\ \bibnamefont {Pfister}},\ }\bibfield  {title} {\enquote {\bibinfo
  {title} {Hidden multipartite entanglement in {{Gaussian}} cluster states},}\
  }\href@noop {} {\bibfield  {journal} {\bibinfo  {journal} {arXiv:1912.06463}\
  } (\bibinfo {year} {2020})}\BibitemShut {NoStop}%
\bibitem [{\citenamefont {Braunstein}(2005)}]{braunstein2005}%
  \BibitemOpen
  \bibfield  {author} {\bibinfo {author} {\bibfnamefont {Samuel~L.}\
  \bibnamefont {Braunstein}},\ }\bibfield  {title} {\enquote {\bibinfo {title}
  {Squeezing as an irreducible resource},}\ }\href {\doibase
  10.1103/PhysRevA.71.055801} {\bibfield  {journal} {\bibinfo  {journal} {Phys.
  Rev. A}\ }\textbf {\bibinfo {volume} {71}},\ \bibinfo {pages} {055801}
  (\bibinfo {year} {2005})}\BibitemShut {NoStop}%
\bibitem [{\citenamefont {Cariolaro}\ and\ \citenamefont
  {Pierobon}(2016{\natexlab{a}})}]{cariolaro2016a}%
  \BibitemOpen
  \bibfield  {author} {\bibinfo {author} {\bibfnamefont {Gianfranco}\
  \bibnamefont {Cariolaro}}\ and\ \bibinfo {author} {\bibfnamefont
  {Gianfranco}\ \bibnamefont {Pierobon}},\ }\bibfield  {title} {\enquote
  {\bibinfo {title} {Reexamination of {{Bloch}}-{{Messiah}} reduction},}\
  }\href {\doibase 10.1103/PhysRevA.93.062115} {\bibfield  {journal} {\bibinfo
  {journal} {Phys. Rev. A}\ }\textbf {\bibinfo {volume} {93}},\ \bibinfo
  {pages} {062115} (\bibinfo {year} {2016}{\natexlab{a}})}\BibitemShut
  {NoStop}%
\bibitem [{\citenamefont {Cariolaro}\ and\ \citenamefont
  {Pierobon}(2016{\natexlab{b}})}]{cariolaro2016}%
  \BibitemOpen
  \bibfield  {author} {\bibinfo {author} {\bibfnamefont {Gianfranco}\
  \bibnamefont {Cariolaro}}\ and\ \bibinfo {author} {\bibfnamefont
  {Gianfranco}\ \bibnamefont {Pierobon}},\ }\bibfield  {title} {\enquote
  {\bibinfo {title} {Bloch-{{Messiah}} reduction of {{Gaussian}} unitaries by
  {{Takagi}} factorization},}\ }\href {\doibase 10.1103/PhysRevA.94.062109}
  {\bibfield  {journal} {\bibinfo  {journal} {Phys. Rev. A}\ }\textbf {\bibinfo
  {volume} {94}},\ \bibinfo {pages} {062109} (\bibinfo {year}
  {2016}{\natexlab{b}})}\BibitemShut {NoStop}%
\bibitem [{\citenamefont {{van Loock}}\ \emph {et~al.}(2007)\citenamefont {{van
  Loock}}, \citenamefont {Weedbrook},\ and\ \citenamefont {Gu}}]{vanloock2007}%
  \BibitemOpen
  \bibfield  {author} {\bibinfo {author} {\bibfnamefont {Peter}\ \bibnamefont
  {{van Loock}}}, \bibinfo {author} {\bibfnamefont {Christian}\ \bibnamefont
  {Weedbrook}}, \ and\ \bibinfo {author} {\bibfnamefont {Mile}\ \bibnamefont
  {Gu}},\ }\bibfield  {title} {\enquote {\bibinfo {title} {Building
  {{Gaussian}} cluster states by linear optics},}\ }\href {\doibase
  10.1103/PhysRevA.76.032321} {\bibfield  {journal} {\bibinfo  {journal} {Phys.
  Rev. A}\ }\textbf {\bibinfo {volume} {76}},\ \bibinfo {pages} {032321}
  (\bibinfo {year} {2007})}\BibitemShut {NoStop}%
\bibitem [{\citenamefont {Gu}\ \emph {et~al.}(2009)\citenamefont {Gu},
  \citenamefont {Weedbrook}, \citenamefont {Menicucci}, \citenamefont {Ralph},\
  and\ \citenamefont {{van Loock}}}]{gu2009}%
  \BibitemOpen
  \bibfield  {author} {\bibinfo {author} {\bibfnamefont {Mile}\ \bibnamefont
  {Gu}}, \bibinfo {author} {\bibfnamefont {Christian}\ \bibnamefont
  {Weedbrook}}, \bibinfo {author} {\bibfnamefont {Nicolas~C.}\ \bibnamefont
  {Menicucci}}, \bibinfo {author} {\bibfnamefont {Timothy~C.}\ \bibnamefont
  {Ralph}}, \ and\ \bibinfo {author} {\bibfnamefont {Peter}\ \bibnamefont {{van
  Loock}}},\ }\bibfield  {title} {\enquote {\bibinfo {title} {Quantum computing
  with continuous-variable clusters},}\ }\href {\doibase
  10.1103/PhysRevA.79.062318} {\bibfield  {journal} {\bibinfo  {journal} {Phys.
  Rev. A}\ }\textbf {\bibinfo {volume} {79}},\ \bibinfo {pages} {062318}
  (\bibinfo {year} {2009})}\BibitemShut {NoStop}%
\bibitem [{\citenamefont {Ferrini}\ \emph {et~al.}(2015)\citenamefont
  {Ferrini}, \citenamefont {Roslund}, \citenamefont {Arzani}, \citenamefont
  {Cai}, \citenamefont {Fabre},\ and\ \citenamefont {Treps}}]{ferrini2015}%
  \BibitemOpen
  \bibfield  {author} {\bibinfo {author} {\bibfnamefont {G.}~\bibnamefont
  {Ferrini}}, \bibinfo {author} {\bibfnamefont {J.}~\bibnamefont {Roslund}},
  \bibinfo {author} {\bibfnamefont {F.}~\bibnamefont {Arzani}}, \bibinfo
  {author} {\bibfnamefont {Y.}~\bibnamefont {Cai}}, \bibinfo {author}
  {\bibfnamefont {C.}~\bibnamefont {Fabre}}, \ and\ \bibinfo {author}
  {\bibfnamefont {N.}~\bibnamefont {Treps}},\ }\bibfield  {title} {\enquote
  {\bibinfo {title} {Optimization of networks for measurement-based quantum
  computation},}\ }\href {\doibase 10.1103/PhysRevA.91.032314} {\bibfield
  {journal} {\bibinfo  {journal} {Phys. Rev. A}\ }\textbf {\bibinfo {volume}
  {91}},\ \bibinfo {pages} {032314} (\bibinfo {year} {2015})}\BibitemShut
  {NoStop}%
\bibitem [{\citenamefont {Zaidi}\ \emph {et~al.}(2008)\citenamefont {Zaidi},
  \citenamefont {Menicucci}, \citenamefont {Flammia}, \citenamefont {Bloomer},
  \citenamefont {Pysher},\ and\ \citenamefont {Pfister}}]{zaidi2008}%
  \BibitemOpen
  \bibfield  {author} {\bibinfo {author} {\bibfnamefont {H.}~\bibnamefont
  {Zaidi}}, \bibinfo {author} {\bibfnamefont {N.~C.}\ \bibnamefont
  {Menicucci}}, \bibinfo {author} {\bibfnamefont {S.~T.}\ \bibnamefont
  {Flammia}}, \bibinfo {author} {\bibfnamefont {R.}~\bibnamefont {Bloomer}},
  \bibinfo {author} {\bibfnamefont {M.}~\bibnamefont {Pysher}}, \ and\ \bibinfo
  {author} {\bibfnamefont {O.}~\bibnamefont {Pfister}},\ }\bibfield  {title}
  {\enquote {\bibinfo {title} {Entangling the optical frequency comb:
  {{Simultaneous}} generation of multiple 2 * 2 and 2 * 3 continuous-variable
  cluster states in a single optical parametric oscillator},}\ }\href {\doibase
  10.1134/S1054660X08050186} {\bibfield  {journal} {\bibinfo  {journal} {Laser
  Phys.}\ }\textbf {\bibinfo {volume} {18}},\ \bibinfo {pages} {659--666}
  (\bibinfo {year} {2008})}\BibitemShut {NoStop}%
\bibitem [{\citenamefont {Flammia}\ \emph {et~al.}(2009)\citenamefont
  {Flammia}, \citenamefont {Menicucci},\ and\ \citenamefont
  {Pfister}}]{flammia2009}%
  \BibitemOpen
  \bibfield  {author} {\bibinfo {author} {\bibfnamefont {Steven~T.}\
  \bibnamefont {Flammia}}, \bibinfo {author} {\bibfnamefont {Nicolas~C.}\
  \bibnamefont {Menicucci}}, \ and\ \bibinfo {author} {\bibfnamefont {Oliver}\
  \bibnamefont {Pfister}},\ }\bibfield  {title} {\enquote {\bibinfo {title}
  {The optical frequency comb as a one-way quantum computer},}\ }\href
  {\doibase 10.1088/0953-4075/42/11/114009} {\bibfield  {journal} {\bibinfo
  {journal} {J. Phys. B: At. Mol. Opt. Phys.}\ }\textbf {\bibinfo {volume}
  {42}},\ \bibinfo {pages} {114009} (\bibinfo {year} {2009})}\BibitemShut
  {NoStop}%
\bibitem [{\citenamefont {Menicucci}(2011)}]{menicucci2011a}%
  \BibitemOpen
  \bibfield  {author} {\bibinfo {author} {\bibfnamefont {Nicolas~C.}\
  \bibnamefont {Menicucci}},\ }\bibfield  {title} {\enquote {\bibinfo {title}
  {Temporal-mode continuous-variable cluster states using linear optics},}\
  }\href {\doibase 10.1103/PhysRevA.83.062314} {\bibfield  {journal} {\bibinfo
  {journal} {Phys. Rev. A}\ }\textbf {\bibinfo {volume} {83}},\ \bibinfo
  {pages} {062314} (\bibinfo {year} {2011})}\BibitemShut {NoStop}%
\bibitem [{\citenamefont {Wang}\ \emph {et~al.}(2014)\citenamefont {Wang},
  \citenamefont {Chen}, \citenamefont {Menicucci},\ and\ \citenamefont
  {Pfister}}]{wang2014a}%
  \BibitemOpen
  \bibfield  {author} {\bibinfo {author} {\bibfnamefont {Pei}\ \bibnamefont
  {Wang}}, \bibinfo {author} {\bibfnamefont {Moran}\ \bibnamefont {Chen}},
  \bibinfo {author} {\bibfnamefont {Nicolas~C.}\ \bibnamefont {Menicucci}}, \
  and\ \bibinfo {author} {\bibfnamefont {Olivier}\ \bibnamefont {Pfister}},\
  }\bibfield  {title} {\enquote {\bibinfo {title} {Weaving quantum optical
  frequency combs into continuous-variable hypercubic cluster states},}\ }\href
  {\doibase 10.1103/PhysRevA.90.032325} {\bibfield  {journal} {\bibinfo
  {journal} {Phys. Rev. A}\ }\textbf {\bibinfo {volume} {90}},\ \bibinfo
  {pages} {032325} (\bibinfo {year} {2014})}\BibitemShut {NoStop}%
\bibitem [{\citenamefont {Alexander}\ \emph {et~al.}(2016)\citenamefont
  {Alexander}, \citenamefont {Wang}, \citenamefont {Sridhar}, \citenamefont
  {Chen}, \citenamefont {Pfister},\ and\ \citenamefont
  {Menicucci}}]{alexander2016b}%
  \BibitemOpen
  \bibfield  {author} {\bibinfo {author} {\bibfnamefont {Rafael~N.}\
  \bibnamefont {Alexander}}, \bibinfo {author} {\bibfnamefont {Pei}\
  \bibnamefont {Wang}}, \bibinfo {author} {\bibfnamefont {Niranjan}\
  \bibnamefont {Sridhar}}, \bibinfo {author} {\bibfnamefont {Moran}\
  \bibnamefont {Chen}}, \bibinfo {author} {\bibfnamefont {Olivier}\
  \bibnamefont {Pfister}}, \ and\ \bibinfo {author} {\bibfnamefont
  {Nicolas~C.}\ \bibnamefont {Menicucci}},\ }\bibfield  {title} {\enquote
  {\bibinfo {title} {One-way quantum computing with arbitrarily large
  time-frequency continuous-variable cluster states from a single optical
  parametric oscillator},}\ }\href {\doibase 10.1103/PhysRevA.94.032327}
  {\bibfield  {journal} {\bibinfo  {journal} {Phys. Rev. A}\ }\textbf {\bibinfo
  {volume} {94}},\ \bibinfo {pages} {032327} (\bibinfo {year}
  {2016})}\BibitemShut {NoStop}%
\end{thebibliography}
%

\end{document}